\magnification=1000
\newcount\magnification
\input  lanlmac.tex
\newskip\defaultbaselineskip\defaultbaselineskip=12pt

\def\gsim{\mathrel{\raise2pt\hbox to 8pt{\raise -5pt\hbox{$\sim$}\hss{$>$}}}}
\def\rsim{\mathrel{\raise2pt\hbox to 8pt{\raise -5pt\hbox{$\sim$}\hss{$>$}}}}
\def\lsim{\mathrel{\raise2pt\hbox to 8pt{\raise -5pt\hbox{$\sim$}\hss{$<$}}}}
\def\ssqr#1#2{{\vbox{\hrule height.#2pt
      \hbox{\vrule width.#2pt height#1pt \kern#1pt\vrule width.#2pt}
      \hrule height.#2pt}\kern-.#2pt}}

\def\listrefsnomod{\footatend\vfill\supereject\immediate\closeout\rfile\writestoppt
\centerline{{\bf References}}\bigskip{\frenchspacing%
\parindent=20pt\escapechar=` \input \jobname.refs\vfill\eject}}
\def\listrefsmag{\footatend\immediate\closeout\rfile\writestoppt
\centerline{{\bf References}}\bigskip{\frenchspacing%
\parindent=20pt\escapechar=` \input \jobname.refs\vfill\eject}}
\ifx\hyperref\undefined\def\hyperref#1#2#3#4{#4}\fi
\ifx\hyperdef\undefined\def\hyperdef#1#2#3#4{#4}\fi
\ifx\href\undefined\def\href#1#2{#2}\fi
\global\newcount\mtabno
\mtabno=1
\def\table#1#2#3{\DefWarn#1%
\xdef #1{\noexpand\hyperref{}{table}{\the\mtabno}%
{\the\mtabno}}\goodbreak\midinsert
$$#2$$\nobreak\centerline{Table~\hyperdef\hypernoname{table}%
{\the\mtabno}{\the\mtabno}. {\sl #3}}\bigskip\endinsert
\writedef{#1\leftbracket#1}\global\advance\mtabno by1}
\def\figure#1#2#3{\DefWarn#1\xdef#1{\noexpand\hyperref{}{figure}%
{\the\figno}{\the\figno}}\writedef{#1\leftbracket#1}%
\figinsert\figin{\centerline{#2}}\medskip\centerline{\vbox{\baselineskip
\defaultbaselineskip
\advance\hsize by -1truein\noindent\wrlabeL{#1=#1}\centerline{\sl
{\bf Fig.~\hyperdef\hypernoname{figure}{\the\figno}{\the\figno}:} #3}}}
\bigskip\endinsert\global\advance\figno by1}
\newread\myread
\def\ignore#1{}%
\def\readdefs{\ifx\writedef\ignore
 \immediate\openin\myread=\jobname.defs
 \ifeof\myread\message{No file \jobname.defs yet.}\else
   \closein\myread\relax\input\jobname.defs\fi\else
 \errmessage{can't \string\readdefs\space after \string\writedefs!!!}
\fi\relax}
\def\makelatexlike#1{\expandafter\let\csname old \string#1\endcsname=#1%
                 \def#1##1{%
                  \edef\tempname{\ifcat\relax\noexpand##1\noexpand##1\else
                                 \expandafter\noexpand\csname##1\endcsname\fi}%
                  \expandafter\expandafter\csname old \string#1\endcsname
                  \tempname}}
\def\nameuse#1{\edef\tempname{\ifcat\relax\noexpand#1\noexpand#1\else
                              \expandafter\noexpand\csname#1\endcsname\fi}%
               \expandafter\ifx\tempname\undefined
                  \message{YET UNDEFINED NAME \string`\string#1\string' USED.}%
                  ???%
               \else\expandafter\tempname\fi}
                                          

\newwrite\figsfile \newwrite\tabsfile
\newif\ifinplacealso\inplacealsofalse
\def\forjournal{
\defaultbaselineskip=24pt
\immediate\openout\figsfile\jobname.figs
\immediate\openout\tabsfile\jobname.tabs
\let\oldfigure=\figure
\let\oldtable=\table
\let\oldend=\end
\def\fourthoffive##1##2##3##4##5{##4}
\def\fourthoffour##1##2##3##4{##4}
\def\getno##1{\expandafter\fourthoffive##1}
\def\ignore##1{}
\ifinplacealso
\def\figure##1##2##3{{\toks0={{\figno=\getno##1\let##1=\undefined
  \let\wrlabeL\ignore\let\writedef\ignore
  \let\hyperref\fourthoffour\let\hyperdef\fourthoffour
  \oldfigure##1{##2}{##3}}\vfill\eject}\immediate\write\figsfile{\the\toks0}%
  \oldfigure##1{##2}{##3}}}
\def\table##1##2##3{{\toks0={{\mtabno=\getno##1\let##1=\undefined
  \let\wrlabeL\ignore\let\writedef\ignore
  \let\hyperref\fourthoffour\let\hyperdef\fourthoffour
  \oldtable##1{##2}{##3}}\vfill\eject}\immediate\write\tabsfile{\the\toks0}%
  \oldtable##1{##2}{##3}}}
\else
\def\figure##1##2##3{{\toks0={{%
  \oldfigure##1{##2}{##3}}\vfill\eject}\immediate\write\figsfile{\the\toks0}}}
\def\table##1##2##3{{\toks0={{%
  \oldtable##1{##2}{##3}}\vfill\eject}\immediate\write\tabsfile{\the\toks0}}}
\fi
\def\end{\vfill\eject\nopagenumbers
         \immediate\closeout\figsfile\input\jobname.figs
         \immediate\closeout\tabsfile\input\jobname.tabs
         \oldend}
}







     \def\CT{{\cal T}}


\def\author{\bigskip\centerline{David Daniel} 
    \smallskip\centerline{\it T-8, MS-B285,
                     Los Alamos National Laboratory, Los Alamos, NM 87545}}



\def\PRL#1{{\it Phys. Rev. Lett.} {\bf #1}}

\def\PRB#1{{\it Phys. Rev.} {\bf B#1}}

\def\etal{{\it et al.\ }}



\input  epsf
\readdefs\writedefs


\ifx\href\undefined\def\href#1#2{{#2}}\fi
\def\spireshome%
{http://www.slac.stanford.edu/cgi-bin/spiface/find/hep/www?FORMAT=WWW&}
{\catcode`\%=12
\xdef\spiresjournal#1#2#3{\noexpand\href{\spireshome
                          rawcmd=find+journal+#1%2C+#2%2C+#3}}
\xdef\spireseprint#1#2{\noexpand\href{\spireshome rawcmd=find+eprint+#1%2F#2}}
\xdef\spiresreport#1{\noexpand\href{\spireshome rawcmd=find+rept+#1}}
}


\def\TAB{\hbox{ $\CT_{\alpha\beta}$}}
\def\TABn{\hbox{ $\CT^n_{\alpha\beta}$}}

\baselineskip 12 pt

\Title{LA UR-96-93}{
CRITICAL EXPONENTS OF THE 3-D ISING MODEL
}
\vskip 0.2 truein

\centerline{Rajan Gupta* $^{\S}$}
\vfootnote*{Invited talk presented at the US-Japan Bilateral Seminar, Maui, August 28-31, 1995. To 
be published in International Journal of Modern Physics C.} 
\vfootnote{\S}{rajan@qcd.lanl.gov}
\smallskip\centerline{\it
  T-8, MS-B285, Los Alamos National Laboratory, Los Alamos, NM 87545}

{\bigskip\centerline{Pablo Tamayo $^\dagger$}
\vfootnote{\dag}{tamayo@think.com.}
\smallskip\centerline{\it
  T-8, MS-B285, Los Alamos National Laboratory, Los Alamos, NM 87545}
 \smallskip\centerline{\it and}
 \smallskip\centerline{\it Thinking Machines Corporation, Cambridge, MA 02143, USA}
}

\vskip 0.2 truein

\centerline{\bf ABSTRACT}
\vskip 0.2 truein

We present a status report on the ongoing analysis of the 3D Ising
model with nearest-neighbor interactions using the Monte Carlo
Renormalization Group (MCRG) and finite size scaling (FSS) methods on
$64^3$, $128^3$, and $256^3$ simple cubic lattices.  Our MCRG
estimates are $K_{nn}^c=0.221655(1)(1)$ and $\nu=0.625(1)$. The FSS
results for $K^c$ are consistent with those from MCRG but the value of
$\nu$ is not. Our best estimate $\eta = 0.025(6)$ covers the spread in
the MCRG and FSS values. A surprise of our calculation is the estimate
$\omega \approx 0.7$ for the correction-to-scaling exponent.  We also
present results for the renormalized coupling $g_R$ along the MCRG
flow and argue that the data support the validity of hyperscaling for
the 3D Ising model.

\vskip 0.2 truein
\Date{Date = 1/14/96 \hfill}

\vfil\eject
\baselineskip 12 pt

\newsec{Introduction}

The 3D Ising model has, over the last 25 years, been used to test the
accuracy of various analytical and numerical methods for solving
Statistical Mechanics systems. In 1992 we presented results of 
simulations on $64^3$ and $128^3$ lattices using the Monte Carlo
Renormalization Group (MCRG) method
\ref\risingus{C.~Baillie, R.~Gupta, K.~Hawick, and S.~Pawley, \PRB{45} (1992) 10438}.
While that work improved on previous MCRG estimates
\ref\Edin{G. S. Pawley, R. H. Swendsen, D. J. Wallace and K. G. Wilson, 
          \PRB{29} (1984) 4030.}\ 
\ref\Blote{H. W. J. Bl\"ote, A. Compagner, J. H. Croockewit,
Y. T. J. C. Fonk, J. R. Heringa, A. Hoogland, T. S. Smit and A. L. van
Willigen, {\it Physica} {\bf A161} (1989) 1.}), it left us with four
unanswered questions. The first and most tantalizing was $--$ are the
exponents rational numbers, $i.e.$ $\nu=0.625$ and $\eta=0.025$.
Second, a more precise determination of the corrections-to-scaling
exponent is needed as it is the largest source of systematic errors.
Third, we wanted to resolve/understand the differences between our
MCRG and existing finite size scaling/ $\epsilon$-expansion results.
Lastly, we wanted to investigate whether hyperscaling holds for this
model.  This talk is a summary of the current status of our
calculations.

In order to address these issues we have extended the calculations in
the following ways. We have made higher statistics runs on $64^3$,
$128^3$, and $256^3$ lattices at $K = 0.221652$ and $0.221655$. On
these lattices we have evaluated, in addition to correlation functions
needed for MCRG studies, quantities needed for finite size scaling
(FSS) analysis and the calculation of the renormalized coupling $g_R$.
As a result we have better estimates of the critical coupling
$K^c_{nn}$, the exponents $\nu$ and $\eta$ from both MCRG and
finite-size scaling (including histogram re-weighting) analysis, and
can address the issue of hyperscaling violations.  Our 
new estimate of the corrections-to-scaling exponent $\omega \approx
0.7$ is significantly smaller than that from other methods. 

All the simulations have been done on the Thinking Machines CM-5 (at
the ACL at LANL) and CM-5E (at TMC) computers.  We used the
Swendsen-Wang cluster update algorithm 
\ref\rSwW{R. Swendsen and J. S. Wang, \PRL{58} (1987) 86.}\ 
and a 250-long 64-bit wide shift-register (Kirkpatrick-Stoll) random
number generator in each vector unit.  The new results agree with our
previous calculation and those in \Edin\ and \Blote. Each of these
calculations used a different random number generator, so their
consistency suggests that there is no obvious bias in the sequence of
random numbers generated (see P. Coddington's talk on random number
generators at this workshop).  Our most extensive results are at
$K_{nn} = 0.221655$, which is our present best estimate of $K_c$,
and the statistical sample consists of 600K, 500K, and 400K
measurements on $64^3$, $128^3$, and $256^3$ lattices respectively.

The details of our implementation of the MCRG method are the same as
in \risingus.  The only change is that we have added 3 more even
couplings (for a total of 56) and one more odd couplings (total 47).
The original 53 even and 46 odd couplings were contained in either a
$3 \times 3$ square or a $2^3$ template \risingus. The new couplings are
those obtained by adding a fourth spin along the cartesian axis to the
$3 \times 3$ template. 

We store the magnetization and energy for each configuration, from
which we can calculate quantities like the specific heat,
susceptibility, Binder's cumulant $U = 3 - \langle m^4 \rangle / \langle
m^2 \rangle^2 $, $etc.$. These results are then evaluated as a
function of $K$, in a small neighborhood of $K_{simulation}$, using the
histogram re-weighting method
\ref\rFaS{A. M. Ferrenberg and R. H. Swendsen, \PRL{61} (1988) 2635.}.
The finite size analysis of these quantities follows the work of
Ferrenberg and Landau
\ref\rFaL{A. Ferrenberg, and D. Landau, \PRB{44} (1991) 5081.}, 
$i.e.$ without corrections to scaling terms. 
To calculate $g_R$, we also need the 
finite lattice correlation length $\xi$.  This is calculated in two ways:
\eqn\exidef{\eqalign{
\langle \sum_{x,y} s(x,y,z) \sum_{x,y} s(x,y,0) \rangle 
    \quad &{\buildrel {\longrightarrow} \over {\scriptstyle z \to \infty } } \quad
     a e^{-z/\xi} ,\cr
{1 \over k^2 } \bigg( {S(0)^2 \over S(k)^2} - 1 \bigg) \ &= \  \xi^2 \,, \cr
}}
where $S(k) = \sum_{x,y,z} s(x,y,z) e^{i \vec k \cdot \vec x}$. We investigate
the 5 lowest momenta but present results only for the lowest, $k_z =
2\pi/L$, as it has the best signal.  With $\xi $ in hand we calculate
$g_R$ defined as
\ref\rbaker{B.~Freedman and G.~Baker, {\it J. Phys. A: Math. Gen.} {\bf 15} (1982) L715.}
\eqn\egrdef{
g_R(K,L) \ = \ \big({L \over \xi} \big)^d \ %
\bigg( 3 -  { \langle m^4 \rangle \over \langle m^2 \rangle^2} \bigg) \ .
}
This is expected to scale as 
\eqn\egrscale{
g_R(K,L) \ \sim \ L^{-w^*}
}
where ${w^*} = (\gamma + d\nu -2\Delta)/\nu$ is referred to as the
anomalous dimension of the vacuum. If hyperscaling holds then $g_R^* \to $
{\it finite non-zero constant} as $L \to \infty$ and $K \to K^*$. 

For the purpose of error analysis the data have been divided into bins
of size 10,000 measurements.  All errors are then calculated by a
single elimination jackknife procedure over these bins. This talk is
organized as follows. We first summarize the MCRG results, then
compare them with FSS estimates, and finally give data for $g_R$, both at 
$K_{sim} = 0.221655$ and along the MCRG flow, and discuss 
hyperscaling.

\newsec{Nearest-Neighbor Critical Coupling $K^c_{nn}$}

We calculate $K^c_{nn}$ using the two lattice $MCRG$ method.  The
results for the two starting couplings $K_{nn} = 0.221652$ and
$0.221655$ and for the two pairs of lattices, $256^3$ versus $128^3$
and $128^3$ versus $64^3$, are shown in Table~\nameuse\tkcmcrg, where
for ease of comparison we have also reproduced results from \risingus.
We expect the deviations $K^c(\infty) - K^c(n)$, where $n$ is the
blocking step, to converge by the geometric factor $\lambda_u /
\lambda_t$ as a function of the blocking level $n$.  For the 3D Ising
model ${\lambda_u / \lambda_t} \approx 6$, and our data for $K^c_{nn}$
(see Table~\nameuse\tkcmcrg) do roughly show convergence by this
factor.

For both starting couplings the data show that the new estimates from
the $128^3/64^3$ lattices analysis are consistent with our earlier
results and give $K^c_{nn} = 0.221652$.  However, the $256^3/128^3$
results show better convergence with respect to blocking steps, and
also show a systematic shift at all blocking levels compared to the
$128^3/64^3$ data. Our new estimate from $256^3/128^3$ comparison is
$K^c_{nn} = 0.221655 \pm 0.000001 $ where only the statistical error
has been quoted.  This shift in $K^c_{nn}$ with lattice size shows
that there could still be finite size corrections at the level of
statistical errors, $i.e.$ a systematic error of $+0.000001$.  So our
final best estimate is
\eqn\ekcnew{
K^c_{nn} = 0.221655 \pm 0.000001 
{\scriptstyle +0.000001 \atop \scriptstyle -0\hphantom{.000001} } \ ,
}
which is consistent with the recent result $0.2216546(10)$ obtained
using FSS analysis by Bl\"ote \etal
\ref\rBlote{H.~Bl\"ote, E.~Luijten, and J.~Heringa, cond-mat/9509016.}. 
In view of this we take the results at $K_{nn} = 0.221655$ to
represent the critical point values.

\table\tkcmcrg{
\vbox{\hbox{\indent\vbox{\tabskip=0pt\offinterlineskip
\def\tlr{\noalign{\hrule}}
\def\tone{0.2216550}
\def\ttwo{0.2216540}
\def\tthr{0.2216520}
\def\tfor{0.2216440}
\def\Lone{256^3/128^3}
\def\Ltwo{128^3/64^3}
\def\myskip{\omit&height2pt& && && && && && && &\cr}
\halign {\strut#& \vrule#\tabskip=0.5em plus2em&
  \hfil$#$\hfil&\vrule#&
  \hfil$#$\hfil&\vrule#&
  \hfil$#$\hfil&\vrule\vrule#&
  \hfil$#$\hfil&\vrule#&
  \hfil$#$\hfil&\vrule#&
  \hfil$#$\hfil&\vrule#&
  \hfil$#$\hfil&\vrule#\tabskip=0pt\cr\tlr
&& Level &&  \tone  &&  \tthr  &&  \tone  &&  \ttwo  &&  \tthr  &&  \tfor  &\cr
\myskip
&& n/m   &&  \Lone  &&  \Lone  &&  \Ltwo  &&  \Ltwo  &&  \Ltwo  &&  \Ltwo  &\cr
\myskip
&&       &&  New    &&  New    &&  New    &&\risingus&&  New    &&\risingus&\cr
\myskip\tlr\tlr
&&  2/1 && 217 \pm 11 && 184 \pm 15 && 095 \pm 13  && 070 \pm 16 && 046 \pm 23  && 095 \pm 17  &\cr\tlr
&&  3/2 && 469 \pm  9 && 458 \pm 18 && 413 \pm 15  && 406 \pm 18 && 394 \pm 27  && 417 \pm 22  &\cr\tlr
&&  4/3 && 537 \pm 10 && 534 \pm 21 && 504 \pm 16  && 500 \pm 21 && 501 \pm 31  && 508 \pm 24  &\cr\tlr
&&  5/4 && 549 \pm 10 && 549 \pm 22 && 523 \pm 16  && 514 \pm 26 && 516 \pm 37  &&             &\cr\tlr
&&  6/5 && 547 \pm 10 &&            &&             &&            &&             &&             &\cr\tlr
}}}}
 
 }
{\vtop{\advance\hsize by -2\parindent 
\noindent
Estimates of $K^c_{nn}$ as a function of the blocking level and
$K_{simulation}$. For brevity only the last three decimal places have
been quoted, so $547\pm10$ is short for $0.2216547\pm0.0000010$. We
have included our old data from \risingus\ to facilitate comparison.
The highest blocked lattice is $4^3$ except at $K_{nn} = 0.221652$
where it is only $8^3$ on $128^3$ and $256^3$ lattices.  The quoted
errors are the statistical errors after averaging the data over the 56
even operators.  The data show a systematic shift between the
$256^3/128^3$ and $128^3/64^3$ lattices.  }}

\newsec{Correlation length exponent $\nu$}

The correlation length exponent $\nu$ is 
determined from the leading even eigenvalue $\lambda_t$ of the linearized 
transformation matrix \TABn, 
\eqn\definenu{
y_t\ \equiv \ {1 \over \nu } \ = \ { { \ln \lambda_t } \over { \ln b } } \ \ ,
}
where $b=2$ is the scale factor of the majority rule blocking
transformation.  Our preferred data for $\lambda_t$
($K_{sim}=0.221655$) is shown in Table~\nameuse\tnuetadata\ as a function
of the blocking step and lattice size.  There are three possible
sources of systematic errors that affect the $L \to \infty$ and $K \to
K_c$ estimates for $\lambda_t$.  These are
\item{1.}
The number of operators measured, $i.e.$ the truncation errors in
evaluating eigenvalues from a finite dimensional \TABn. We find that
the number of operators needed to achieve convergence increases with
the blocking level $n$. The data show that with the 56 even operators
the eigenvalues show convergence at all levels.  (The same
is true in the sector of odd interactions from which we extract the exponent
$\eta$).  Even at the highest blocked level there is no detectable
variation after including 30 operators.  Unfortunately, the
convergence with the number of operators is not monotonic and there is 
no independent way of confirming that the results have converged. Thus, 
we cannot estimate the possible error due to lack of convergence, 
and guess that it is smaller than the statistical error.

\item{2.} Finite size effects on blocked lattices. It has been observed
in \risingus, \Edin, and \Blote\ that finite size effects are
discernible only when blocking from $8^3 \to 4^3$ lattices or smaller.
The correction increases the estimate of $\lambda_t$. In Ref.~\Blote\
the correction in $\lambda_t^{8 \to 4}$ was estimated to be $0.02$.
Our estimate based on comparing $256^3, 128^3, 64^3$ lattices is
$\approx 0.01$.  Applying $+0.01$ as the correction to our $256^3$
lattices data, we get the lower limit $\lambda_t^{8 \to 4} = 3.008$
corresponding to $\nu = 0.6294$. We discuss the $L \to \infty$ limit
below.

\item{3.}
Error in the estimate of $K^c_{nn}$. The value of $\lambda_t$ also
increases with $K^c_{nn}$ as shown by the data in Fig.~\nameuse\fnu, 
and in Ref.~\Edin. The dependence of $\lambda_t$ on 
$K^c_{nn}$ is marginal on the first couple of blocking levels and then
increases rapidly with $n$ for $n > 3$. Since our estimate of
$K^c_{nn}$ is converging from below, our results at $K=0.221655$ may
underestimate $\lambda_t$.

\noindent The bottom line is that the systematic effects discussed in 
items 2 and 3 will tend to increase $\lambda_t$ or equivalently decrease $\nu$.

\table\tnuetadata{
\vbox{\hbox{\indent\vbox{\tabskip=0pt\offinterlineskip
\def\tlr{\noalign{\hrule}}
\def\tone{0.2216550}
\def\ttwo{0.2216540}
\def\tthr{0.2216520}
\def\tfor{0.2216440}
\def\Lone{256^3}
\def\Ltwo{128^3}
\def\Lthr{64^3}
\def\myskip{\omit&height2pt& && && && && && && &\cr}
\def\mybskip{\omit&height2pt& && \multispan{5} && \multispan{5} &\cr}
\halign {\strut#& \vrule#\tabskip=0.5em plus2em&
  \hfil$#$\hfil&\vrule\vrule#&
  \hfil$#$\hfil&\vrule#&
  \hfil$#$\hfil&\vrule#&
  \hfil$#$\hfil&\vrule\vrule#&
  \hfil$#$\hfil&\vrule#&
  \hfil$#$\hfil&\vrule#&
  \hfil$#$\hfil&\vrule#\tabskip=0pt\cr\tlr
\mybskip
&&     && \multispan{5} {\hfil $\lambda_t$ \hfil} && \multispan{5}{\hfil $\lambda_h$ \hfil} &\cr
\mybskip\tlr
&&     &&  \Lone  &&  \Ltwo  &&  \Lthr &&  \Lone  &&  \Ltwo  &&  \Lthr  &\cr
\myskip\tlr\tlr
&& 0/1 && 2.681(2) &&  2.685(2) && 2.684(3) && 5.4948(06) && 5.4941(06) && 5.4948(08)&\cr
&& 1/2 && 2.843(2) &&  2.847(2) && 2.847(2) && 5.5050(02) && 5.5052(02) && 5.5050(03)&\cr
&& 2/3 && 2.930(3) &&  2.930(3) && 2.930(3) && 5.5501(02) && 5.5499(03) && 5.5494(07)&\cr
&& 3/4 && 2.969(3) &&  2.973(4) && 2.971(5) && 5.5741(04) && 5.5745(08) && 5.5701(21)&\cr
&& 4/5 && 2.995(5) &&  2.985(7) &&          && 5.5845(11) && 5.5826(31) &&           &\cr
&& 5/6 && 2.998(7) &&           &&          && 5.5850(31) &&            &&           &\cr\tlr
}}}}
 
 }
{\vtop{\advance\hsize by -2\parindent 
\noindent
Estimates of $\lambda_t$ and $\lambda_h$ as a function of the blocking level and 
lattice size for $K=0.221655$. 
}}

\figure\fnu{\line{\hss\epsfysize=2.94in\epsfbox{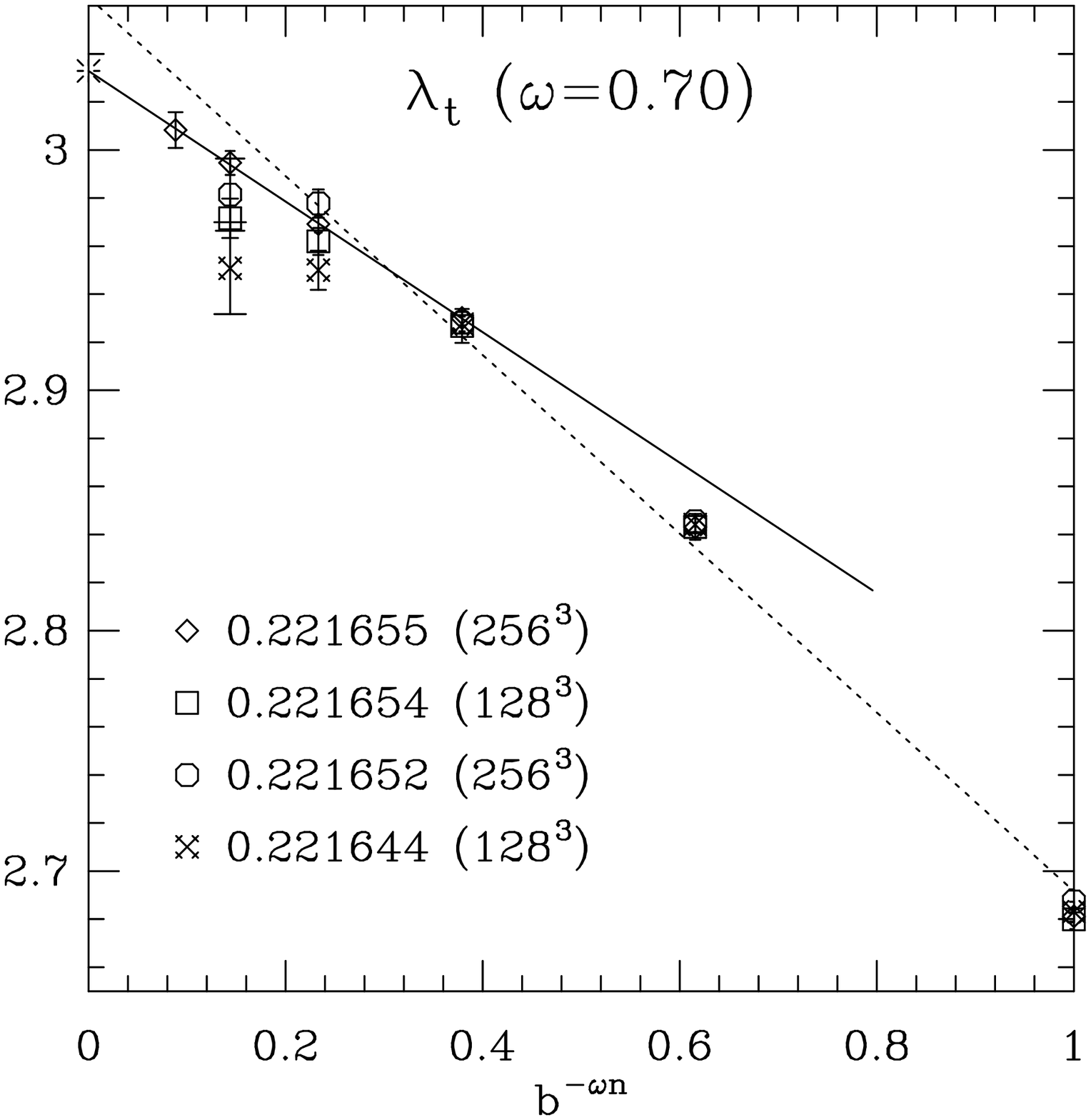}\hss\epsfysize=2.9in\epsfbox{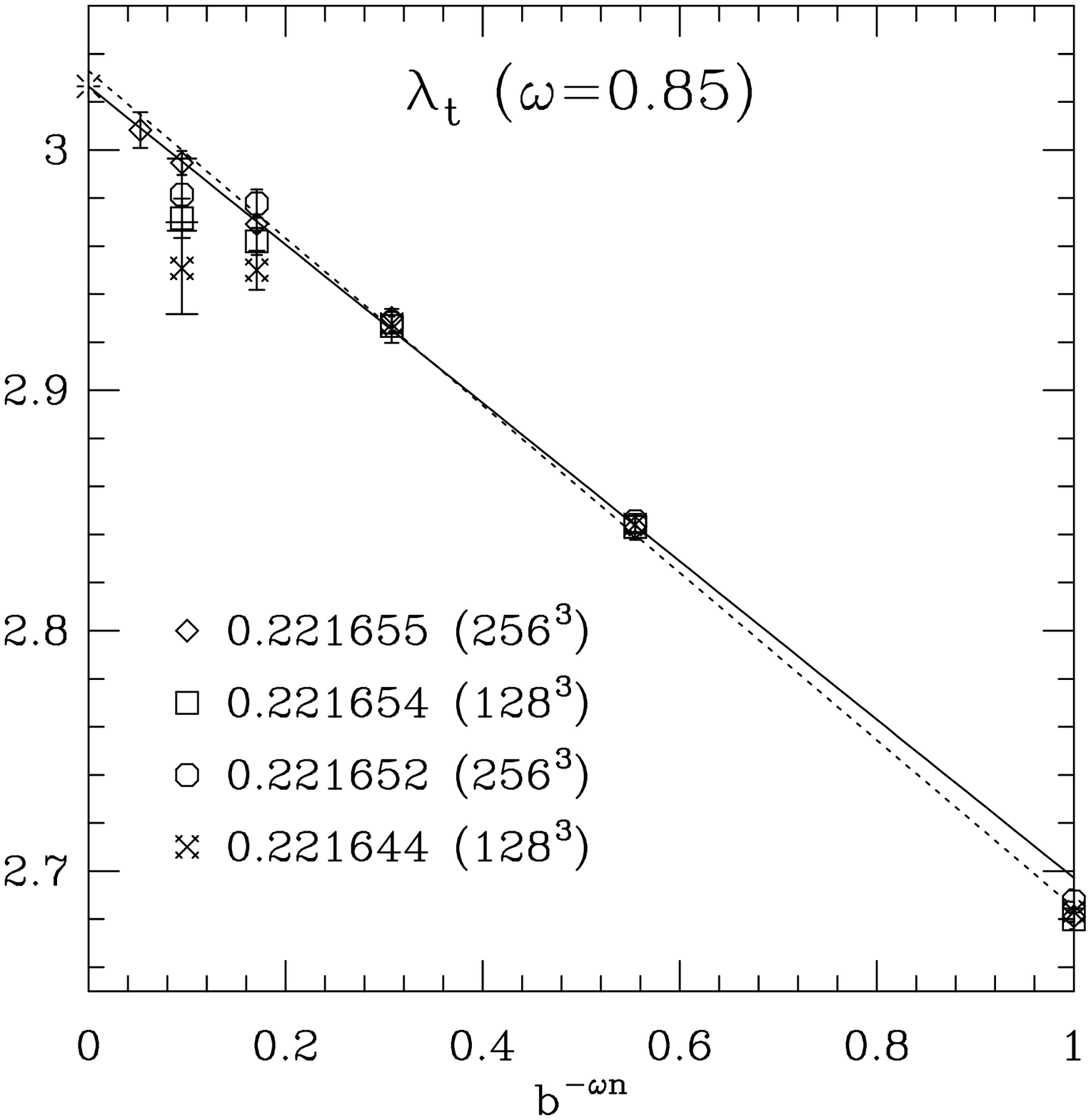}\hss}}
{\vtop{\advance\hsize by -2\parindent \noindent 
Estimates of $\lambda_t$ as a function of the blocking level $n$ for
simulations at different couplings $K$. The best estimate of the
extrapolated value with $\omega=0.70$ ($\omega=0.85$) is from the linear fit
to the diamonds skipping $n=0,1$ ($n=0$) points. The fit to all the
diamond points is, for comparison, shown by the dotted line.}}


Finally, we are interested in the value of $\lambda_t$ at the 
fixed point.  To obtain this we extrapolate $\lambda_t$ versus the 
blocking level $n$ using \risingus
\eqn\ytextrapolation{
\lambda_t (n) \ = \ \lambda_t^* \ + \ a_t b^{- \omega n} \ .
} 
where $\omega = \theta/\nu$ is the leading correction-to-scaling
exponent.  There are two issues that need to be addressed in doing
this extrapolation in the number of blocking steps $n$ ($i.e.$ the $L
\to \infty$ limit).  The first is the value of $\omega$ and the second
is whether the fit should exclude the first few blocking steps to
avoid transients, to account for which requires further corrections to
the leading behavior shown in Eq.\ytextrapolation.  The
calculation of $\omega$ is discussed in the next section and our
present estimate $\omega = 0.7$ is surprisingly low.  We, therefore
present an analysis for $\omega = 0.7$ and $0.85$, where the second
estimate is roughly what is given by other methods (FSS,
$\epsilon$-expansion, $etc$. See
\ref\fisher{S. Zinn and M. Fisher, Maryland Preprint, Nov 1995.}\ for a very
recent survey).  The question of transients is completely empirical,
$i.e.$ we neglect data at initial blocking steps until $\chi^2 \sim
1$.

On basis of the quality of the fit to the $K=0.221655$, $L=256$
data the best estimates for the two extreme values of $\omega$ are
\eqn\resultnu{\eqalign{
\lambda_t^* \ = \ 3.028(3) \quad \Longrightarrow \quad \nu \ = \ 0.6256(\hphantom{1}5)  
                \qquad\qquad (\omega=0.85,\ n=2-6) \ , \cr
\lambda_t^* \ = \ 3.033(6) \quad \Longrightarrow \quad \nu \ = \ 0.6247(10)  
                \qquad\qquad (\omega=0.70,\ n=3-6) \ . \cr
}}
These two estimates are consistent, we therefore take the mean value
and the larger of the two errors to get our present best estimate $\nu
= 0.625(1)$. To improve this result will require a better estimate of
$\omega$ and data on larger lattices (more blocking steps). 

\newsec{Correlation function exponent $\eta$}

The correlation function exponent $\eta$ is given by 
\eqn\defineeta{
\eta \ = \ d+2 -2 { {\log \lambda_h} \over \log b}\ \equiv \ d+2 -2 y_h \ \ ,
}
where $b=2$, $d=3$ and $\lambda_h$ is the largest eigenvalue of \TAB\
constructed from the odd interactions.  The discussion of the type and
sign of the various systematic errors in the extraction of $\lambda_h$
is the same as for $\lambda_t$. The raw data are shown in
Table~\nameuse\tnuetadata, and the value of finite size correction we
apply to $\lambda_h^{8 \to 4}$ is $0.002$.  Then, from the $L=256$
data ($\lambda_h^{8 \to 4}=5.587(3)$) we get the upper bound
$\eta=0.0359(16)$.


\figure\feta{\line{\hss\epsfysize=3.0in\epsfbox{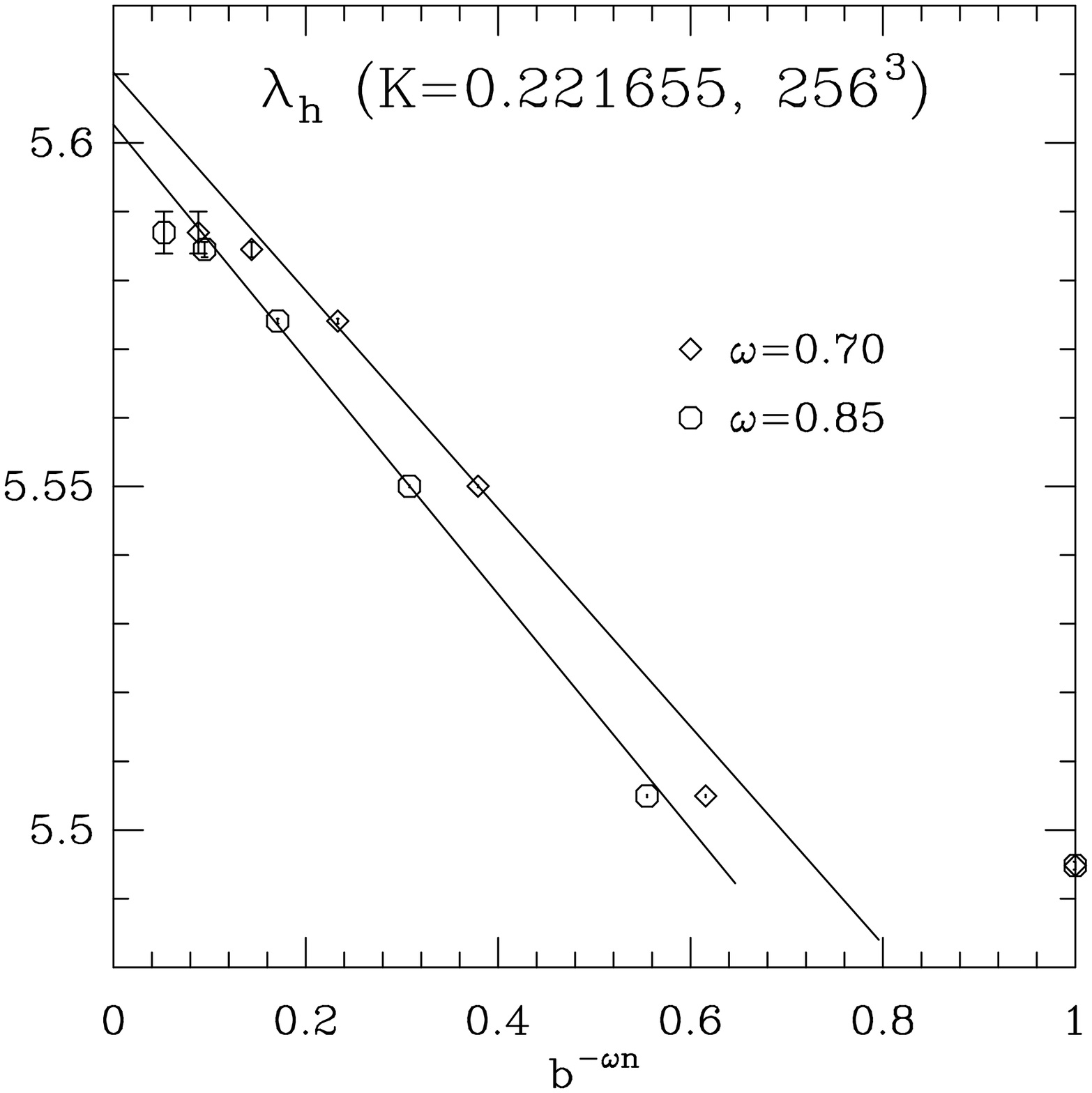}\hss\epsfysize=3.0in\epsfbox{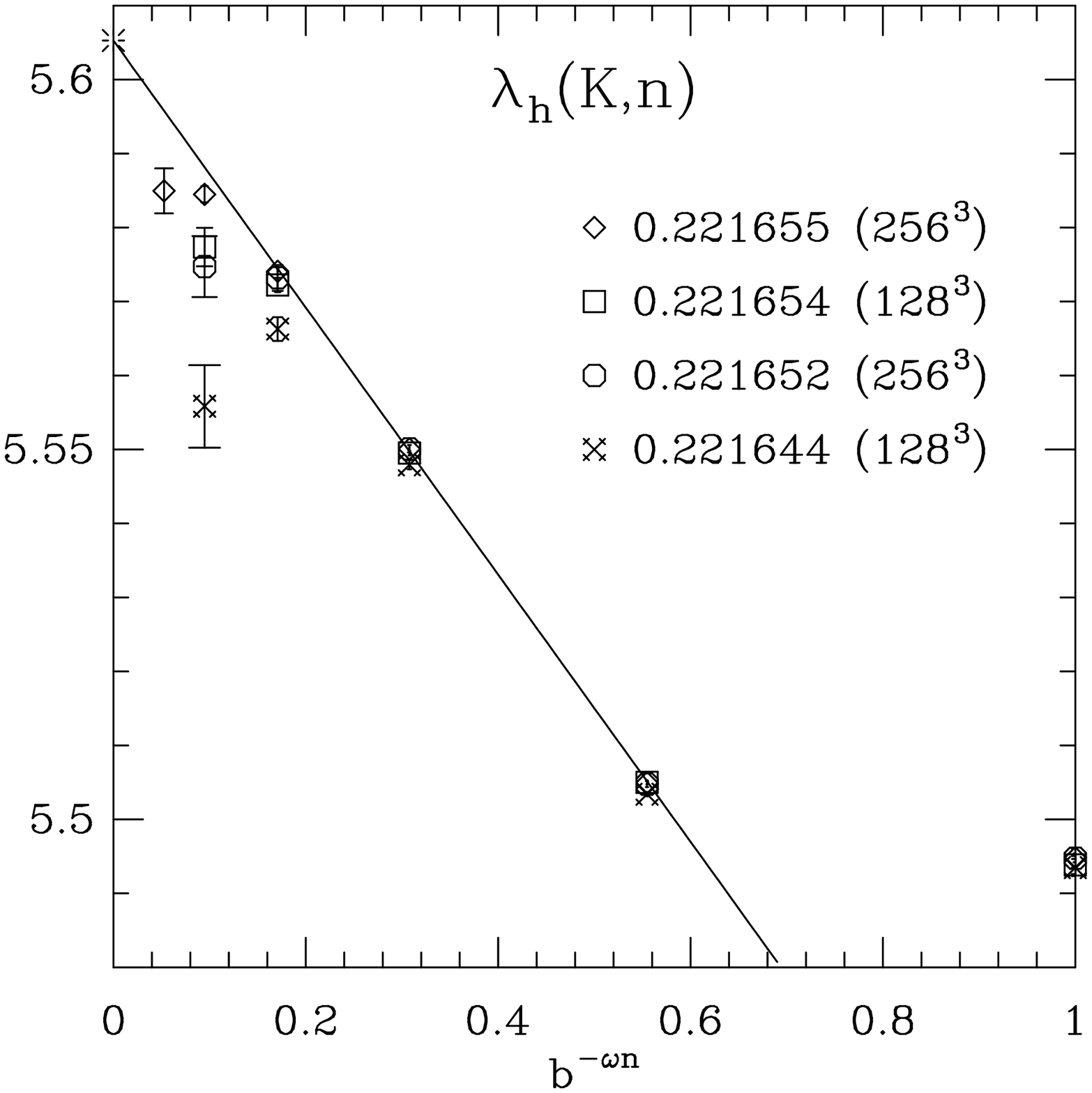}\hss}}
{\vtop{\advance\hsize by -2\parindent 
\noindent (A) Estimates of
$\lambda_h$ as a function of the blocking level $n$ for simulations on
$256^3$ lattices at $K=0.221655$. The same data is plotted for
$\omega=0.70$ and $0.85$, and for each case we show a linear fit to $n=2-5$ points. 
The $n=5$ point deviates significantly from the fits. (B) $\lambda_h$ as a function of $K$ and $n$.}}

To extrapolate to $L \to \infty$ we proceed in exactly the same way as
for $\lambda_t$. However, as exemplified by Fig.~\nameuse\feta, the
points at $n=4,5$ show significant deviations from the linear fits.
Even though the fit with $\omega=0.85$ is somewhat better, the $n=5$
point raises questions about the validity of the linear extrapolation.
There are two possibilities. One, the value flattens out at
$\lambda_h=5.59$, in which case $\eta = 0.034$.  Second, the points at
higher blocking levels are not well determined (note the large
dependence of $\lambda_h$ on $K$ at higher $n$ as shown in
Fig.~\nameuse\feta.  The systematic errors could therefore be larger than
the statistical), and the linear extrapolation is valid. In the latter
case one gets
\eqn\resulteta{\eqalign{
\lambda_h^* \ = \ 5.603(4) \quad \Longrightarrow \quad \eta \ = \ 0.028(2)  
                \qquad\qquad (\omega=0.85,\ n=2-5) \ , \cr
\lambda_h^* \ = \ 5.610(5) \quad \Longrightarrow \quad \eta \ = \ 0.024(3)  
                \qquad\qquad (\omega=0.70,\ n=2-5) \ . \cr
}}
For errors we have used the difference between the
extrapolated values with fits to $n=1-5$ and $n=2-5$ points. The
bottom line is that even though we have improved the estimates for
$\lambda_h$ on lattices of size up to $L=256$, there is still a large
ambiguity in the determination of $\eta$ due to the extrapolation to
$L=\infty$.

\newsec{Corrections-to-scaling exponent $\omega$}

It should be clear from the above discussion that a precise estimate of
$\omega$ is very important in order to take the $L \to \infty$ limit.  In a MCRG
calculation the correction-to-scaling exponent is determined from the
sub-leading eigenvalue $\lambda_{t,2}$ in the even sector; $\omega
\equiv - y_{t,2} = {- \log \lambda_{t,2} / \log b} $.  Only if $\omega$
is known can the exponents $\nu$ and $\eta$, calculated along a
critical RG flow, be extrapolated to the fixed point using Eq.
\ytextrapolation.  We have therefore spent considerable effort in
estimating $\omega$.  With our best data, shown in Table~\nameuse\tomegadata, 
we have overcome the statistical problem that
plagued the data in \risingus. The second and third eigenvalue no
longer merge into a complex pair when the number of operators is $\ge
15$. However, we now find that the value of $\omega$ decreases both with the
number of operators and blocking steps; it starts off at $\approx
0.85$ on the first blocking step and with 10 operators, but finally 
settles down to 
\eqn\omegaresult{
\omega \ \sim \ 0.70   \quad \Longrightarrow \quad \theta = \omega\nu \sim 0.44 \ \ .
}
This value is significantly smaller than the world average,
$\theta=0.54(3)$, of estimates obtained using other methods as
presented in \fisher. Clearly, this issue requires further attention.

\table\tomegadata{
\vbox{\hbox{\indent\vbox{\tabskip=0pt\offinterlineskip
\def\tlr{\noalign{\hrule}}
\def\tone{0.2216550}
\def\ttwo{0.2216540}
\def\tthr{0.2216520}
\def\tfor{0.2216440}
\def\Lone{256^3}
\def\Ltwo{128^3}
\def\Lthr{64^3}
\def\myskip{\omit&height2pt& && && && && && &\cr}
\halign {\strut#& \vrule#\tabskip=0.5em plus2em&
  \hfil$#$\hfil&\vrule#&
  \hfil$#$\hfil&\vrule#&
  \hfil$#$\hfil&\vrule#&
  \hfil$#$\hfil&\vrule#&
  \hfil$#$\hfil&\vrule#&
  \hfil$#$\hfil&\vrule#\tabskip=0pt\cr\tlr
\myskip
&&    &&  10\ Ops &&  20\ Ops && 30\ Ops && 40\ Ops && 50\ Ops  &\cr
\myskip\tlr\tlr
&& 1 && 0.90(3) && 1.01(4) && 0.97(4) && 0.98(4) && 0.97(4) &\cr
&& 2 && 0.85(3) && 0.83(4) && 0.79(4) && 0.80(4) && 0.79(4) &\cr
&& 3 && 0.81(3) && 0.76(4) && 0.73(4) && 0.72(4) && 0.73(4) &\cr
&& 4 && 0.79(3) && 0.74(4) && 0.70(4) && 0.67(4) && 0.67(4) &\cr
&& 5 && 0.75(3) && 0.73(3) && 0.70(3) && 0.70(3) && 0.70(3) &\cr\tlr
\cr}}}}
 
 }
{\vtop{\advance\hsize by -2\parindent 
\noindent
Estimates of $\omega$ as a function of the blocking level and the
number of operators used in constructing the transformation matrix
${\cal T}_{\alpha\beta}$. The data is for $K=0.221655$.  }}

\newsec{Finite size scaling using the histogram re-weighting method.}

We use the histogram re-weighting method \rFaS\ 
to estimate (i) the position of the maximum for various thermodynamic
quantities, (ii) the value at this maximum, and (iii) the value of $K$
at the point of crossing of $U$ and $g_R$ for two different size
lattices. The method consists of building a histogram $H(E,m)$, $i.e.$
the number of configurations with energy $E$ and magnetization $m$,
using an equilibrium (canonical) Monte Carlo simulation at temperature
$K_{sim}$. With this histogram, the equilibrium probability
distribution at other temperatures $K$ is
\eqn\ePKdef{
P_K (E,m) = { H(E,m) \exp [ \Delta K E ] \over \sum_{E,m} H(E,m) \exp [ \Delta K E]}, 
}
where $\Delta K = K - K_{sim}$. The average value of any
function of $E$ and $m$, $Q(E,m)$, at coupling $K$ is then given by
\eqn\eexpatK{
\langle Q_K(E,m) \rangle = \sum_{E,m} Q(E,m) P_K (E,m).
}
The value of thermodynamic derivatives with respect to $K$ are 
obtained from Monte Carlo measurements of correlation functions,
\eqn\edQdKdef{
{d \langle Q \rangle \over d K } = \langle Q E \rangle - 
\langle Q \rangle \langle E \rangle .
}
Again, by using the re-weighting technique these correlation functions
can be evaluated at all temperatures in a certain neighborhood of
$K_{sim}$.  Thus, the location and magnitude of the peaks, and the points 
of crossings can be obtained from simulations at a single temperature.

The propagation of errors under this re-weighting is not
straightforward and has been dealt with by Ferrenberg and by Swendsen
in their talks at this meeting.  In our current analysis the error
estimates are the naive statistical ones and
ignore all correlations and uncertainty in determining $H(E,m)$.
We only present results for $K_{sim} = 0.221655$ as these have higher
statistics and correspond to our estimate of the infinite volume
$K_c$. With this and the re-weighted data generated from it in hand we use
the appropriate finite size scaling relations to derive estimates of
the critical exponents and temperature.

\medskip
\figure\fhnugamma{\line{\hss\epsfysize=2.94in\epsfbox{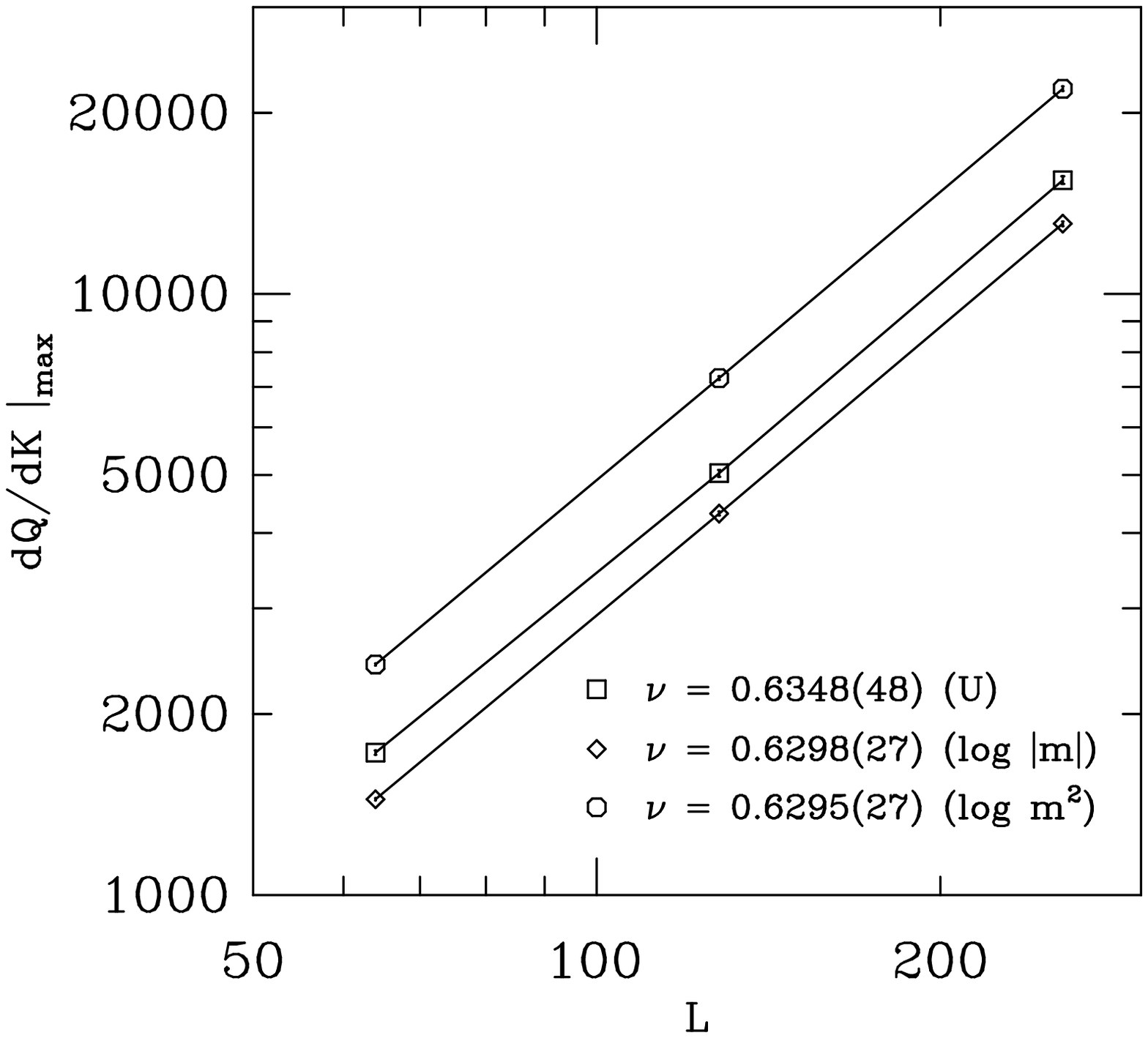}\hss\epsfysize=2.9in\epsfbox{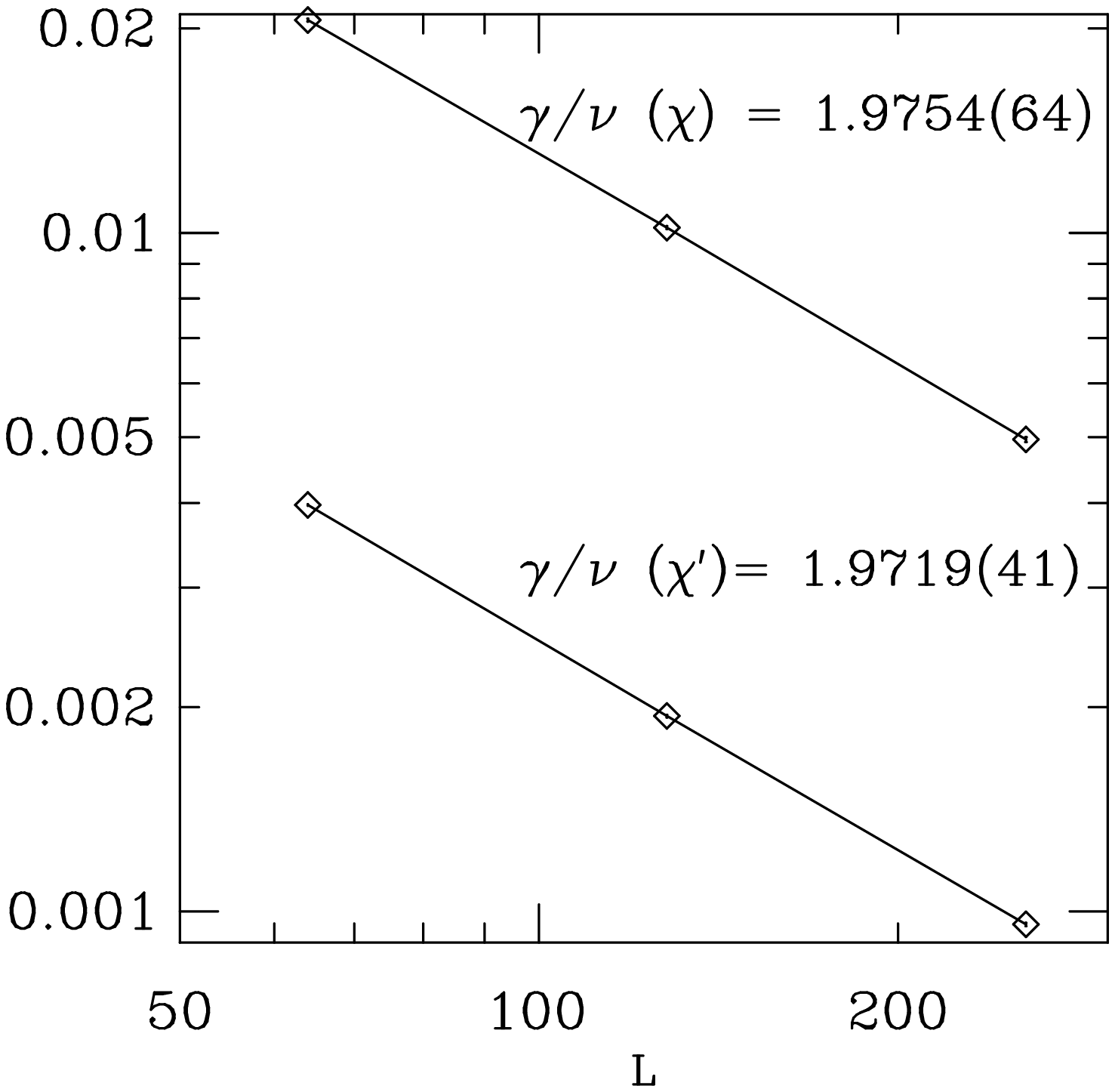}\hss}}
{\vtop{\advance\hsize by -2\parindent \noindent 
(A) Estimates of $\nu$ from the finite size scaling of the maxima of 
the derivatives of $U$, $\log m$ and $\log m^2$ with respect to $K$. 
(B) The slope of $\log \chi$ versus $\log L$ gives the exponent ratio $\gamma/\nu$
assuming the leading order FSS relation $\chi \sim L^{\gamma/\nu}$.}}

\noindent{\bf Estimate of $\nu$}: 
This is obtained from the finite size scaling of the
maxima of thermodynamic derivatives of $U$, $\log |m|$, and $\log m^2$
\rFaL.  For example, ${d \langle \log m^2 \rangle / d K } = 
{\langle m^2 E \rangle / \langle m^2 \rangle} - \langle E \rangle $ is
calculated as a function of $K$ using the histogram $H(E,m)$ to
re-weight the data.  FSS analysis to extract $\nu$ is done keeping
only the leading term in the scaling behavior
\eqn\ehnuscaling{
\left. {d Q \over d K } \right|_{max} \ = \ a L^{1/\nu} 
\big( 1 + b L^{-\omega} + \ldots \big) \ ,\,,
}
as we cannot reliably include correction terms with data at only 3
values of $L$.  Linear fits to the maximum value versus $L^{1/\nu}$
are shown in Fig.~\nameuse\fhnugamma. The quality of the fits is exceedingly 
good, and the final results are
\eqn\ehresult{\eqalign{
 \nu & =  0.6348(48) \;\;\;\; U       \cr
 \nu & =  0.6298(27) \;\;\;\; \log |m| \cr
 \nu & =  0.6295(27) \;\;\;\; \log m^2 \ . \cr
}}
The values obtained from the derivative of $\log |m|$ and $\log m^2$
agree, while that from $U$ is higher by its $ 1 \sigma$
error estimate.  These estimates are higher than the MCRG value by roughly 
one combined $\sigma $. 

\figure\fhkc{\epsfysize=3in\epsfbox{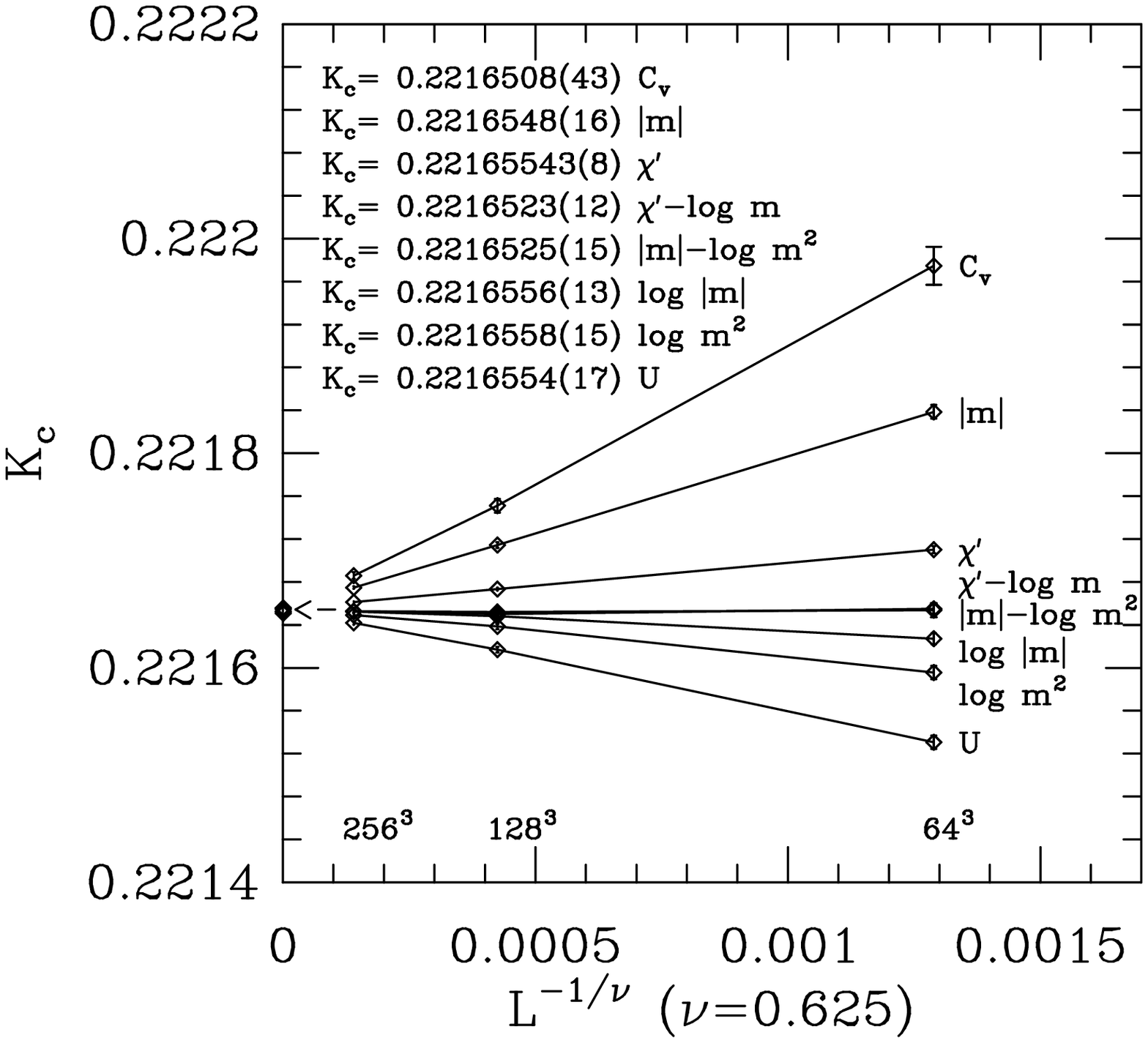}}
{\vtop{\advance\hsize by -2\parindent \noindent 
Estimate of $K_c(L)$ obtained from the FSS analysis of different thermodynamic 
quantities assuming $\nu=0.625$. The data are for $K_{sim}=0.221655$.}}

\medskip
\noindent{\bf Critical Coupling $K_c$}: $K_c$ 
has been calculated in two different ways.
One, we considered the location of the maxima of different thermodynamic
derivatives as a function of system size.  The finite size scaling
behavior of these $K_c(L)$ is \rFaL,
\eqn\eKcscaling{
K_c(L) = K_c + a' L^{-1/\nu} \big( 1 + b' L^{-\omega} + \ldots \big) \ ,
}
where the values of $a'$ and $b'$ are different for each thermodynamic
quantity.  The three data points for each quantity shown in
Fig.~\nameuse\fhkc\ have been joined together by straight lines to
highlight the deviations from linearity.  With just three lattice
sizes we cannot include the correction term ($i.e.\ \omega$), to take
into account the deviations from linearity.  In all quantities, the
magnitude of these deviations is surprisingly large considering that
the data of Ferrenberg and Landau \rFaL\ show a good fit to the linear
form at smaller ($L=24-96$) lattice sizes.  We do not understand this
discrepancy at present other than mention the possibility that in our
$64^3$ data $\Delta K$ may be too large for the re-weighting to be reliable.
Since $\chi'$ and $m$ converge from above while $\log |m|$ and $\log
m^2$ approach it from below, we have constructed linear combinations,
$\chi' - \log |m|$ and $|m| - \log m^2$, that show much smaller
corrections. As pointed out in
\rFaL, the quantity farthest away from $K_c$ is $C_v$ and it gives $K_c$
with the largest errors.

Fig.~\nameuse\fhkc\ was generated using $\nu=0.625$. To determine
$K_c$ we show, in Table~\nameuse\thkc, the variation with $\nu$ of
$K_c$ obtained using the data from $128^3$ and $256^3$ lattices to
extrapolate to $L=\infty$.  We find that there is no significant
variation with $\nu$ in the canonical range of values $0.62-0.635$.
Different observables give $K_c$ in the range $0.221653 - 0.221656$
with a typical statistical error of $0.000002$.  For our best
estimates we take the mean of the various estimates (excluding that
from $C_v$) with $\nu=0.625$. The result is
\eqn\ehKcresult{
K_c \  = \  0.2216544(20)(15)
}
where the first error is statistical and the second is an estimate of
the systematic error based on the variation with observable type. This
value is lower than that obtained by Ferrenberg and Landau \rFaL, but
consistent with our MCRG result.

\table\thkc{
\vbox{\hbox{\indent\vbox{\tabskip=0pt\offinterlineskip
\def\tlr{\noalign{\hrule}}
\halign {\strut#& \vrule#\tabskip=0.5em plus2em&
  \hfil$#$\hfil&\vrule#&
  \hfil$#$\hfil&\vrule#&
  \hfil$#$\hfil&\vrule#&
  \hfil$#$\hfil&\vrule#&
  \hfil$#$\hfil&\vrule#\tabskip=0pt\cr\tlr
&& Observable &&  \nu=0.62  &&  \nu=0.625   &&  \nu=0.63  &&  \nu=0.635  &\cr
\omit&height2pt&
\omit&height2pt&
\omit&height2pt&
\omit&height2pt&
\omit&height2pt&
\omit&\cr\tlr\tlr
&& |m|               &&0.2216559(22)&&0.2216553(22)&&0.2216551(22)&&0.2216548(23)&\cr
&& U                 &&0.2216538(24)&&0.2216539(24)&&0.2216541(24)&&0.2216543(24)&\cr
&& \log |m|          &&0.2216547(17)&&0.2216547(18)&&0.2216548(18)&&0.2216548(18)&\cr
&& \log m^2          &&0.2216545(21)&&0.2216545(21)&&0.2216546(21)&&0.2216547(21)&\cr
&& C_v               &&0.2216544(58)&&0.2216540(58)&&0.2216536(58)&&0.2216531(59)&\cr
&& \chi'             &&0.2216555(10)&&0.2216554(11)&&0.2216553(11)&&0.2216553(11)&\cr
&& \chi'- \log |m|   &&0.2216540(17)&&0.2216540(17)&&0.2216540(18)&&0.2216540(18)&\cr
&& |m| - \log m^2    &&0.2216529(21)&&0.2216529(21)&&0.2216529(21)&&0.2216529(21)&\cr\tlr
}}}}
 
 }
{\vtop{\advance\hsize by -2\parindent 
\noindent
FSS estimates of $K^c_{nn}$ from different observables as a function of 
$\nu$. 
}}

A second estimate of $K_c$ is given by the point of crossing of $U(L)$
and $g_R(L)$ calculated on two lattices of different sizes. Again, the
use of re-weighting trick to extend the data to temperatures in the
vicinity of $K_{sim}$ is essential. Our results are shown in
Fig.~\nameuse\fU. The final values, taken from the
crossing point of $L=128$ and $L=256$ lattices data, are
\eqn\ehKccrossing{\eqalign{
 K_c^{U}   &= 0.2216560  \;\;\;\;   U(K_c) = 1.409(9)  \,, \cr
 K_c^{g_R} &= 0.2216551  \;\;\;\;   g_R(K_c) = 5.23(10) \,. \cr
}}
Note that the error estimates do not take into account the
correlations in the data generated by the re-weighting. It is
interesting to note that the estimate of crossing point obtained from
$L=64$ and $L=128$ lattices is $\sim 0.221652$, indicating a
convergence from below. Also, the estimates of $K_c$ from the two
pairs of lattice sizes are in very good agreement with the
corresponding results obtained from MCRG analysis.

\figure\fU{\line{\hss\epsfysize=2.94in\epsfbox{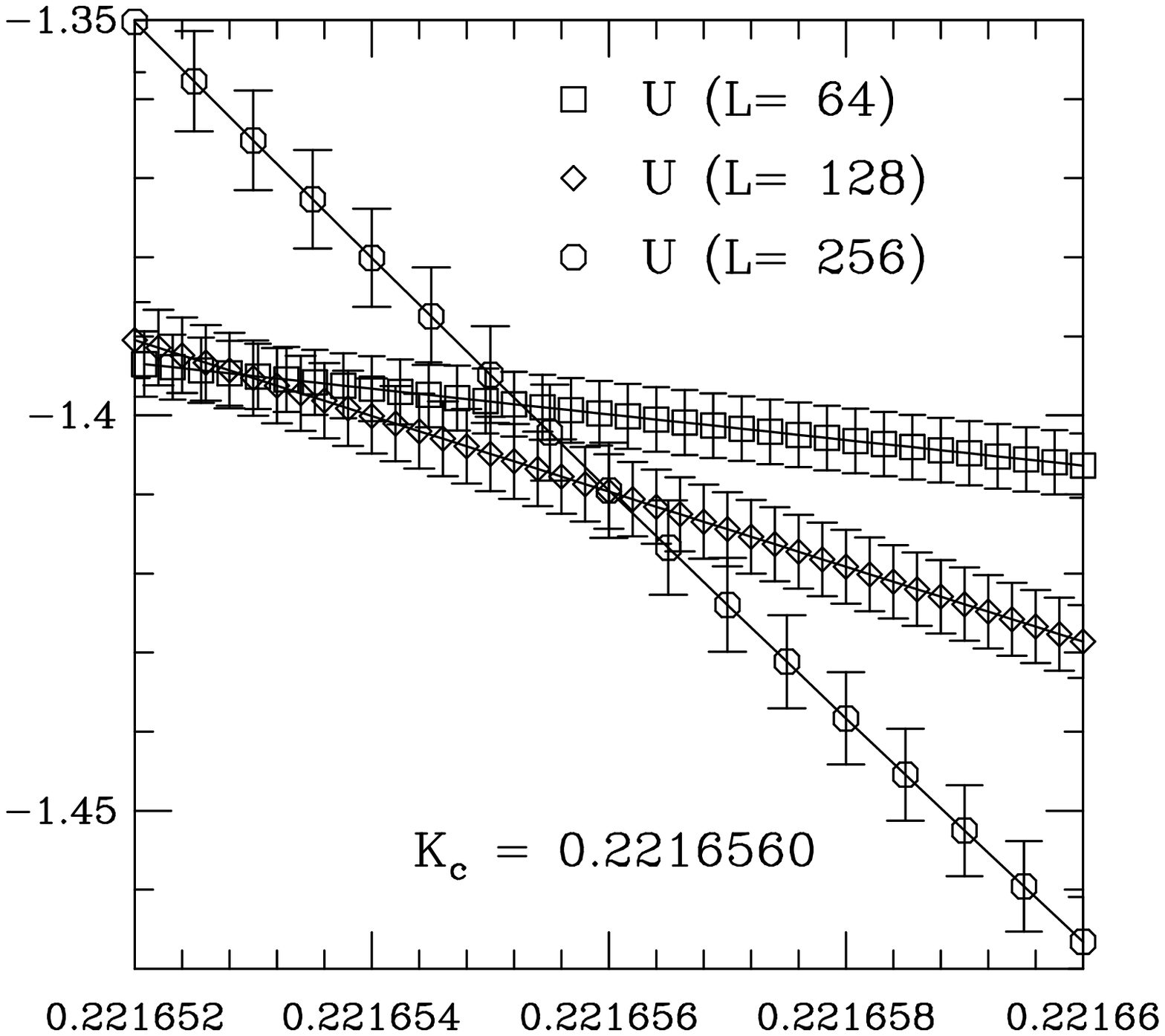}\hss\epsfysize=2.9in\epsfbox{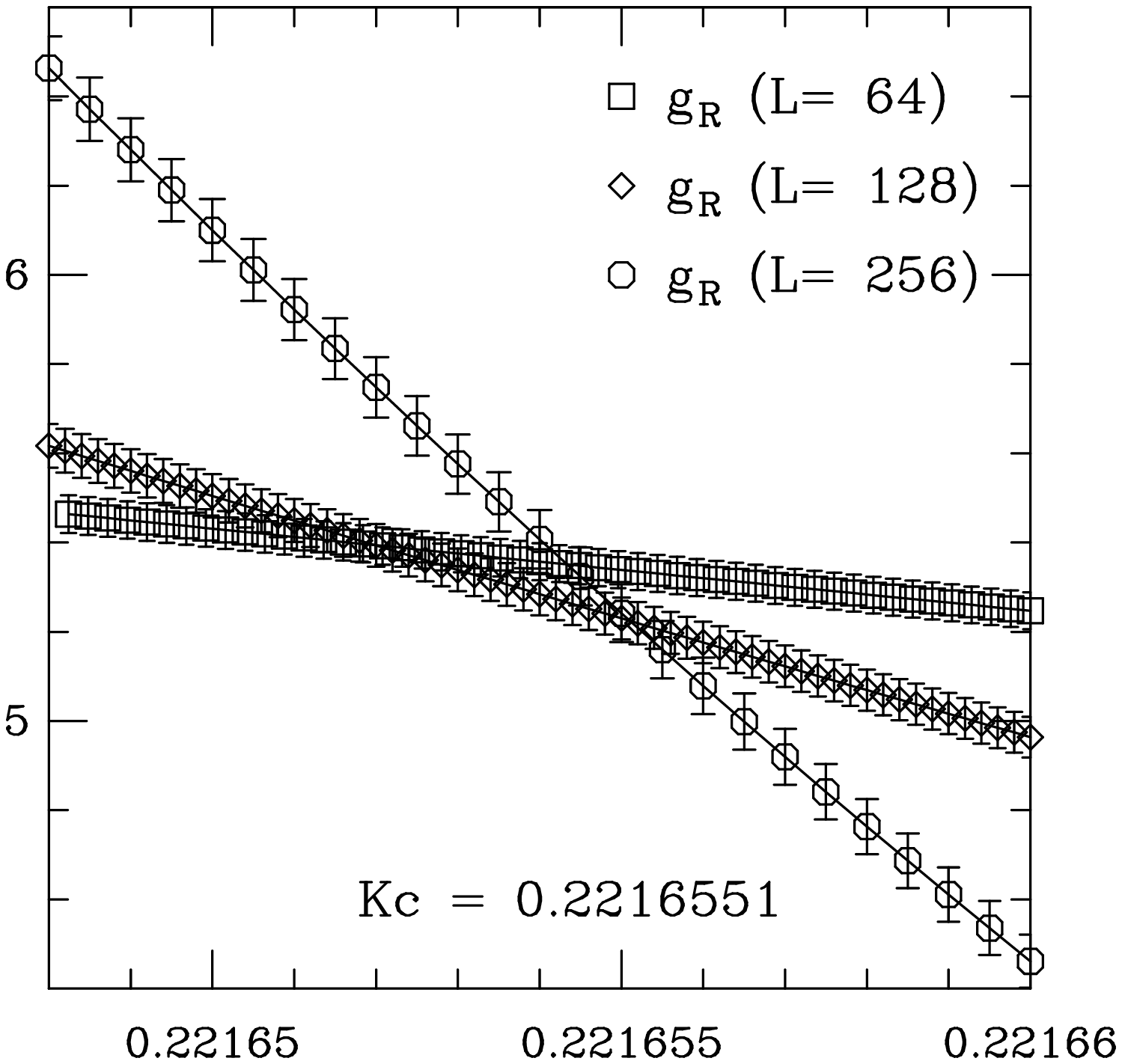}\hss}}
{\vtop{\advance\hsize by -2\parindent \noindent Plot of $U$ and $g_R$
as a function of $K$ obtained from the histogram re-weighting method.
The $K_c(L)$ estimated from the crossing point of $128$ and $256$
lattices data is indicated. Note that the values
and error estimates are highly correlated. }}

\medskip
\noindent{\bf Critical Exponent $\gamma$}: 
We estimate $\gamma$ from the finite size scaling of the
susceptibility. A linear fit to $\log \chi$ versus $\log L$ is shown
in Fig.~\nameuse\fhnugamma\ for data at $K=0.221655$ and for L= $64^3,
128^3$ and $256^3$ lattices. From the slope we obtain
\eqn\ehgammaresult{
  \gamma/\nu  \ =\ 1.9754(64) \,.
}
(To allow detailed comparison with the results in \rFaL\ we also give
the value $ \gamma/\nu = 1.9719(41)$ obtained using $\chi'$.)  Using
the hyperscaling relation $\eta = 2-\gamma/\nu$ we now get $\eta =
0.0246(64)$.  This estimate is consistent with the MCRG result
obtained assuming the validity of the linear approximation. We thus
quote $\eta=0.025(6)$ as our best value since it covers the various
estimates.

\newsec{Renormalized coupling $g_R$ and hyperscaling}

It has been pointed out by Baker and Kawashima 
\ref\baker{G.~Baker and N.~Kawashima, \PRL{75} (1995) 994.}\ 
that the two limits $L \to \infty$ and $K \to K_c$ do not commute and
that there is a discontinuity in the value of $g_R$ when calculated in
the following two ways. Our approach, which is to fix $K =
K^c(\infty)$ and then slowly take the thermodynamic limit versus the
``right'' way which is to always keep $L/\xi$ large and fixed (the
precise value depends on the model and should be representative of the
thermodynamic limit) while taking $K \to K^c$.  Note that the
field-theoretic value $g_R = 23.73(2)$
\ref\bakerbook{G.~Baker, {\it Quantitative Theory of Critical Phenomena}, 
               Academic Press, 1990.}\ is an estimate representative
of the ``right'' order of limits.  Nevertheless, one can establish
hyperscaling by our non-perturbative approach if we can show that
$g_R$ converges to a non-zero lower bound as $L \to \infty$.

The data for $g_R$ in the vicinity of $K^c$ is shown in
Fig.~\nameuse\fU.  One expects an increase in the slope with lattice
size as the discontinuity due to the interchange of limits becomes
sharper with lattice size.  This is borne out by the data. On the
other hand the crossing value shows a small decrease between the
$128^3/64^3$ and $256^3/128^3$ lattices, $i.e.$ the data does not 
converge from below. Based on our data (2 crossing points) we guess
that $g_R(\infty) \approx 5$.  We regard this non-zero result as a
weaker verification of hyperscaling than one would have liked.

Another necessary condition to test whether hyperscaling holds, on basis of 
the data for $g_R$ in Fig.~\nameuse\fU, is whether it 
is representative of the fixed point value. To check this we have also
calculated $g_R$ on the blocked lattices. The data on $256^3 \to 128^3
\to 64^3 \to 32^3$ lattices is virtually identical, indicating that
our estimate does not change along the flow to the fixed point. On
smaller lattices the two methods for calculating $\xi$ give different
results and the use of finite lattice versus continuum energy-momentum
dispersion relation makes a difference. We therefore consider our data
on lattices smaller than $32^3$ unreliable for the purposes of this
test.

\newsec{Conclusions}

Our new results $K_{nn}^c = 0.221655(1)$, $\nu=0.625(1)$, and
$\eta=0.025(6)$ are an improvement over the previous MCRG values.
These values continue to support the notion that the exponents are
rational.  We have also reconciled the disagreement between finite
size scaling and MCRG results for $K_{nn}^c$ and $\eta$ by a
comparative study using the same data. The value of the exponent $\nu$
from the two methods, however, shows a significant difference. One
possible explanation is the lack of various corrections-to-scaling
terms in our FSS analysis.  

The big surprise of the current MCRG analysis is the result $\omega
\approx 0.7$, which is significantly lower than all previous
estimates. This issue clearly needs to be investigated further.

The convergence of $g_R^*$, defined to be the crossing point value in 
the limit $L \to \infty$,
seems to be from above. Thus, our data does not provide the desired
lower bound to validate hyperscaling.  We estimate $g_R^*(L = \infty)$
from data at the two crossing points to be $\sim 5$. If this non-zero
value withstands further scrutiny, then we will have established that
hyperscaling holds for the 3D Ising model.

\bigskip
\bigskip
\centerline{\bf Acknowledgements}
\bigskip

It is a pleasure to thank David Landau and Masuo Suzuki for organizing a
very informative workshop in such idyllic surroundings. We thank George 
Baker and Robert Swendsen for informative discussions. We are also
grateful to the tremendous support provided by the Advanced Computing
Laboratory and Thinking Machines Corporation for this project.


\bigskip
\bigskip
\listrefsmag

\end